\documentclass[runningheads]{llncs}
\usepackage[T1]{fontenc}
\usepackage{lmodern}
\usepackage{amsmath,amssymb}
\usepackage{booktabs}
\usepackage{float}
\usepackage{array}
\usepackage{graphicx}
\usepackage[hyphens]{url}
\usepackage{xcolor}
\usepackage{listings}
\usepackage{tikz}
\usetikzlibrary{arrows.meta,patterns}
\usepackage[hidelinks]{hyperref}
\hypersetup{bookmarksdepth=2,bookmarksopen=true,bookmarksopenlevel=2,
  bookmarksnumbered=true,pdfpagemode=UseOutlines}
\newcommand{\csfs}{\texttt{OP\_CHECKSIGFROMSTACK}}
\newcommand{\ctv}{\texttt{OP\_CHECKTEMPLATEVERIFY}}
\newcommand{\checksig}{\texttt{OP\_CHECKSIG}}

\newcommand{\sha}{\mathsf{SHA256}}

\newif\ifanonymous
\anonymousfalse
\begin{document}
\renewcommand{\thelstlisting}{\arabic{lstlisting}}

\title{CTV Without CTV: Verified Template Checks and
Reusable State Authorization with CSFS+CAT}
\titlerunning{CTV Without CTV with CSFS+CAT}

\ifanonymous
\author{Anonymous Authors}
\authorrunning{Anonymous Authors}
\institute{}
\else
\author{Baixing Liu\inst{1} \and Aaron Zhang\inst{2} \and
Ramble Lan\inst{2} \and Xuan Ding\inst{1}}
\authorrunning{B. Liu, A. Zhang, R. Lan, and X. Ding}
\institute{Tsinghua University\\
\email{liubx25@mails.tsinghua.edu.cn, dingx04@gmail.com}
\and
Unaffiliated\\
\email{aaron@btcdevlab.org, rambledefi@gmail.com}}
\fi

\maketitle

\begin{abstract}
Can template authorization remain reusable when a payment channel state spends a different previous output? We use the proposed CAT and CHECKSIGFROMSTACK (CSFS) opcodes to construct a Tapscript macro that checks the 32-byte BIP-119 CHECKTEMPLATEVERIFY (CTV) predicate for a defined family of single-input Taproot transactions with fixed output-script lengths. A publicly refreshable auxiliary signature, checked by both CSFS and CHECKSIG, binds witness-supplied fields to the actual transaction. A separate participant signature authorizes the template hash and remains valid across eligible previous outputs sharing that hash. Unlike the auxiliary signature, this authorization need not commit to the spent outpoint, amount, or tapleaf. We prove soundness, public completeness, and stack preservation for the macro, and give conditions for substituting it into a CTV+CSFS state-replacement protocol. An executable compiler and Bitcoin Inquisition experiments on regtest and public Signet demonstrate a confirmed stale-state override and delayed settlement in a two-party, three-update prototype with public recovery data. The measured update costs 496 vB, versus 241 vB with native CTV+CSFS. The result is a bounded, higher-cost route to state replacement under proposed opcodes, without native CTV or ANYPREVOUT.
\end{abstract}

\section{Introduction}
\label{sec:intro}

Covenants restrict how an output may be spent, supporting applications such as vaults, transaction trees and payment channels~\cite{moser2016covenants,bartoletti2020unchained,swambo2020threeways}. Payment-channel designs motivate enforceable off-chain state updates and settlement~\cite{decker2015duplex,poon2016lightning}. Sprites formalizes state channels~\cite{miller2019sprites}. Different consensus proposals supply different building blocks. \ctv{} compares a transaction-template hash with a stack value~\cite{bip119}. ANYPREVOUT (APO) changes which transaction fields a signature commits to, allowing a previously issued authorization to be attached to another compatible previous output~\cite{bip118}. The latter capability supports the state-replacement design of eltoo, also called Lightning Symmetry~\cite{decker2018eltoo}.

Concatenation and stack-message signature verification provide another way to inspect a transaction. A script reconstructs a signature message, verifies a signature for its hash with \csfs{}, and verifies the same signature with ordinary \checksig{}. Successful verification binds the reconstruction to the actual transaction. O'Connor and Piekarska demonstrated this same-signature covenant technique in Elements Alpha~\cite{oconnor2017enhancing}; Poelstra studied CAT and Schnorr variants for Bitcoin~\cite{poelstra2021cat1,poelstra2021cat2}. Neither idea is claimed as new here. Our question is whether the proposed BIP-347/348 Tapscript semantics can support a check of every BIP-119 template field, preserve a composable stack interface, and reuse participant authorization across previous outputs, with executable state-replacement evidence. Checking only the output hash or a partial input predicate would answer a narrower question.

BIP-346 proposes direct, selectable transaction hashing~\cite{bip346}, while BIP-348 describes a CTV+CSFS authorization block~\cite{bip348}. Rubin and Rearden analyze rekeying and key ladders for CTV+CSFS state protocols~\cite{rubin2024rekeying}. Our construction instead supplies the template check using CAT+CSFS and then composes it with separate state authorization. It does not improve their cost or establish general equivalence with those proposals.

That last observation is sometimes taken to rule out reusable authorization. The inference is too strong. The signature used to authenticate reconstructed data need not be the signature through which channel participants authorize a state. If the reconstruction uses a dedicated public auxiliary secret, anybody can issue a fresh auxiliary signature for the actual spending transaction. A separate, secret-key signature can continue to authorize the same template hash. This separation is the central mechanism studied here.

\begin{figure}[tb]
\centering
\includegraphics[width=\linewidth]{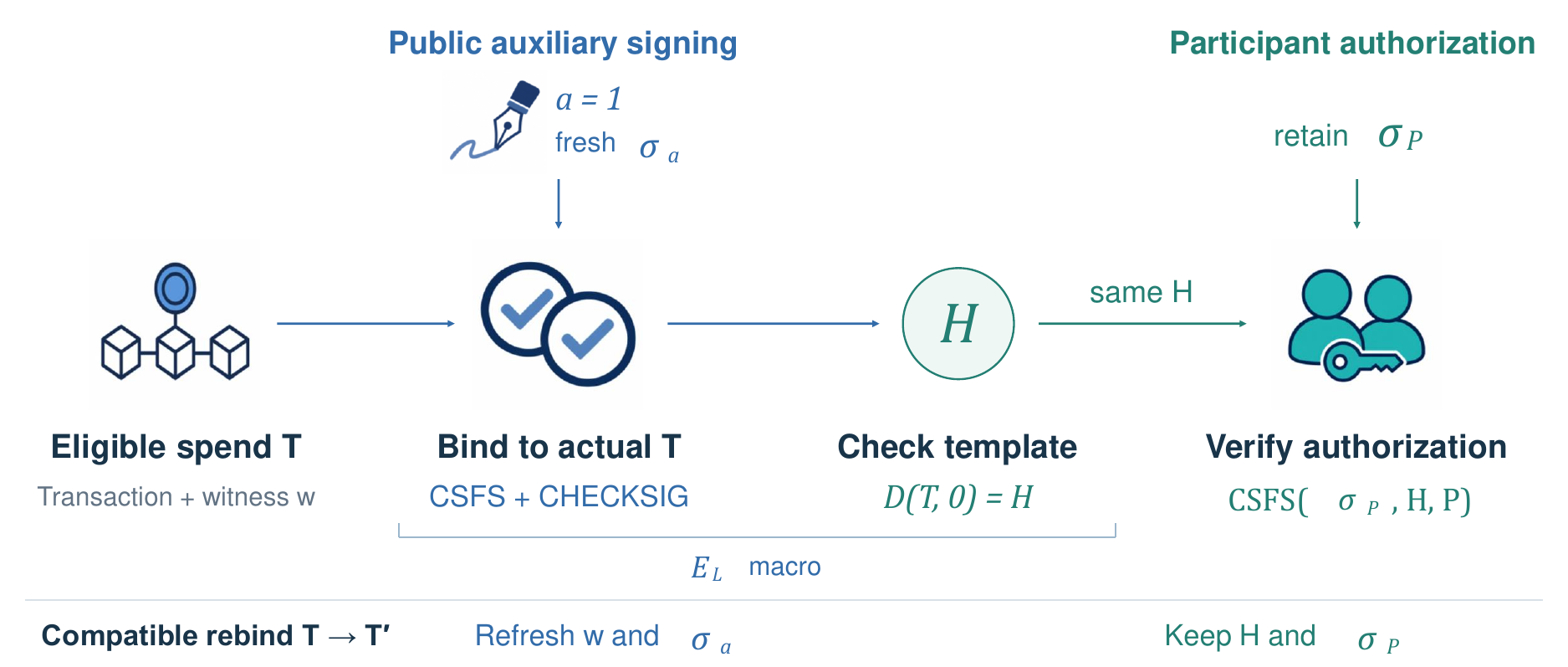}
\caption{Template checking with separate auxiliary and participant signatures. The public helper signs for the actual spend using $a=1$. CSFS and CHECKSIG verify the same auxiliary signature under $Q=G$, binding the reconstruction to $T$. The macro $E_L$ checks $D(T,0)=H$ and preserves $H$ for the participant check. For a compatible rebind to $T'$ with the same template, the public witness is refreshed while $\sigma_P$ is retained. Reuse requires eligible transactions in $\mathcal D_L$ (Section~\ref{sec:family}) and policies accepting the same $(H,\sigma_P,P)$; state replacement also requires Section~\ref{sec:composition}'s conditions.}
\label{fig:core-idea}
\end{figure}

Figure~\ref{fig:core-idea} summarizes the construction. The auxiliary witness binds the reconstructed data to the spending transaction, while the participant signature authorizes the same template hash that the macro checks and preserves. Rebinding to a compatible previous output refreshes the former without requiring a new participant authorization.

We make three contributions. First, we construct and prove a CTV-template check for a precisely specified, script-enforced transaction family, including input and output counts. Second, we give an executable opcode compiler, distinguish auxiliary signing from authorization, and prove an authorization-reuse corollary. Third, we state the conditions under which the check can replace CTV in a compatible protocol and evaluate a bounded state-replacement implementation against native CTV+CSFS on regtest and public Signet. The experiment includes confirmed stale-state override, unchanged participant authorization, public data recovery, and delayed settlement. Our contribution is the scoped construction, its interface and its executable evidence; earlier work already established the underlying sighash-introspection technique.

The paper proceeds from the signature model to message binding, template emulation, protocol composition, and validation. We prove transaction-level statements and a conditional protocol reduction. The Signet experiments exercise the new construction, but we do not claim an unrestricted replacement for every BIP-119 behavior or a complete Lightning implementation.

\section{Background and Scope}
\label{sec:background}

\subsection{Signature messages and script operations}

Let $T$ be a transaction and $i$ its current input. Under SIGHASH\_DEFAULT, script-path Taproot signing with no annex uses a 212-byte preimage $M(T,i)$. This includes the epoch byte, the BIP-341 message and the BIP-342 extension~\cite{bip340,bip341,bip342}. Put $\tau=\sha(\mathtt{TapSighash})$ and
\[
 h(T,i)=\sha(\tau\|\tau\|M(T,i)).
\]
This 32-byte digest is the message supplied to BIP-340 verification. The CSFS proposal verifies a stack-supplied message without adding a further hash or domain tag~\cite{bip348}. The construction must therefore supply exactly $h$, not an independently tagged approximation.

CAT concatenates the second-top stack item followed by the top item. Its result must not exceed 520 bytes~\cite{bip347}. CSFS consumes a signature, a message and a public key, with the public key at the top. Ordinary CHECKSIG obtains the message from the transaction rather than the stack. Both checks must succeed; a nonempty signature alone is not sufficient. We use only valid, fixed 32-byte public keys, avoiding the success behavior assigned to unknown key lengths.

\subsection{The CTV template and the implemented transaction family}
\label{sec:family}

Write $\operatorname{le}_k(x)$ for a $k$-byte little-endian encoding. For an otherwise valid transaction whose input scriptSigs are empty, the BIP-119 default template hash is
\begin{equation}
\begin{split}
D(T,i)=\sha(&v\|\ell\|\operatorname{le}_4(n_{\rm in})\|S\\
            &\|\operatorname{le}_4(n_{\rm out})\|O\|\operatorname{le}_4(i)),
\end{split}
\label{eq:ctv}
\end{equation}
where $v$ and $\ell$ are the four-byte version and locktime encodings, $S$ hashes all four-byte input sequences, and $O$ hashes all serialized outputs. With nonempty scriptSigs, BIP-119 inserts an additional hash; that case is outside our construction~\cite{bip119}. We consider only the 32-byte CTV argument case, not its NOP behavior on other lengths.

Fix at compilation time a nonempty vector $L=(l_1,\ldots,l_m)$ satisfying
\[
0\leq l_j\leq252,\qquad B_L=\sum_{j=1}^{m}(9+l_j)\leq520.
\]
An output of script length $l_j$ has eight value bytes, one CompactSize length byte, and $l_j$ script bytes. Define $\mathcal D_L$ to contain otherwise consensus-valid spends with exactly one native Taproot input, input index zero, no annex, and exactly $m$ outputs of these script lengths. The executing tapscript contains no executed CODESEPARATOR. The script bytes themselves, amounts, version, locktime, sequence, and outpoint may vary. In particular, output scripts need not be P2TR: a profile may include an OP\_RETURN output for protocol data availability.

These conditions define the construction implemented and proved here. They let the script reconstruct and hash the complete output serialization in one stack element and prove both vector counts. The 520-byte bound applies to each CAT result; its application to the entire output vector follows from this construction's single-element reconstruction. It is not an upper bound established here on the transaction sizes that CSFS+CAT can support. Likewise, the single-input and fixed-length profiles delimit this implementation, rather than establish that more general profiles are impossible. Multi-input fee sponsorship, variable output-script lengths, and nonempty scriptSigs are outside the present theorem.

\subsection{Adversary and authorization model}

The adversary controls the witness, transaction fields and witness order. It may know the auxiliary secret $a=1$ and therefore sign any message under the fixed BIP-340 public key $Q=G$. It must not obtain the separate protocol authorization secret. We assume collision and second-preimage resistance for the relevant SHA-256 constructions and resistance to finding distinct BIP-340 messages with the same challenge scalar for fixed nonce commitment and key. Authorization additionally assumes unforgeability of the protocol's signature scheme and its signing policy.

Our statements concern a specified tapleaf. The UTXO must not have an alternative path that bypasses its policy. In particular, the public auxiliary key must \emph{not} be used as an unprotected Taproot internal key: its known secret would permit a key-path spend. A script-only setup can use a suitable NUMS internal point; cooperative key-path spending requires a separate, explicit policy. The local test vectors use the NUMS point described in BIP-341, rather than $Q$, as their internal key.

\section{Message Binding with a Public Auxiliary Secret}
\label{sec:binding}

The witness supplies chunks from which the script computes $R$. It checks
\begin{equation}
\operatorname{CSFS}(\sigma_a,\sha(\tau\|\tau\|R),Q)
\;\land\;
\operatorname{CHECKSIG}(\sigma_a,Q).
\label{eq:binding}
\end{equation}
The same 64-byte signature and the same valid key are used in both arms. A 64-byte CHECKSIG signature selects SIGHASH\_DEFAULT. The compiler rejects other lengths explicitly.

\begin{figure}[tb]
\centering
\begin{tikzpicture}[x=1cm,y=1cm,>=Latex,
  every node/.style={font=\small,align=center},
  box/.style={draw,rounded corners=1pt,minimum height=.65cm,text width=4.15cm,inner sep=3pt},
  flow/.style={->,line width=.55pt}]
\node[font=\small\bfseries] at (2.4,5.05) {Script reconstruction};
\node[font=\small\bfseries] at (8.85,5.05) {Consensus message};
\node[box] (w) at (2.4,4.3) {Witness pieces $w$ and constants $C$};
\node[box] (t) at (8.85,4.3) {Spending transaction $T$, input $i$};
\node[box] (r) at (2.4,3.15) {$R(w,C)$\\Concatenation and field checks};
\node[box] (m) at (8.85,3.15) {$M(T,i)$\\Taproot signature preimage};
\node[box] (rh) at (2.4,2) {$\sha(\tau\|\tau\|R(w,C))$};
\node[box] (mh) at (8.85,2) {$h(T,i)=\sha(\tau\|\tau\|M(T,i))$};
\node[box,text width=2.55cm] (csfs) at (2.4,.85) {CSFS verification};
\node[box,text width=2.55cm] (cs) at (8.85,.85) {CHECKSIG verification};
\node[text width=2.1cm,font=\small] (sig) at (5.625,.85) {Same $(\sigma,P)$};
\node[draw,minimum height=.65cm,text width=5.1cm,fill=black!6] (ok) at (5.625,-.5) {Both checks accept\\$\Longrightarrow\ R(w,C)=M(T,i)$};
\draw[flow] (w)--(r);
\draw[flow] (t)--(m);
\draw[flow] (r)--(rh);
\draw[flow] (m)--(mh);
\draw[flow] (rh)--(csfs);
\draw[flow] (mh)--(cs);
\draw[flow] (sig.west)--(csfs.east);
\draw[flow] (sig.east)--(cs.west);
\draw[flow] (csfs.south)--(2.4,-.5)--(ok.west);
\draw[flow] (cs.south)--(8.85,-.5)--(ok.east);
\end{tikzpicture}
\caption{Same-signature binding of a reconstructed Taproot message to the transaction's consensus message. Both checks use the same auxiliary signature and public key.}
\label{fig:same-signature}
\end{figure}
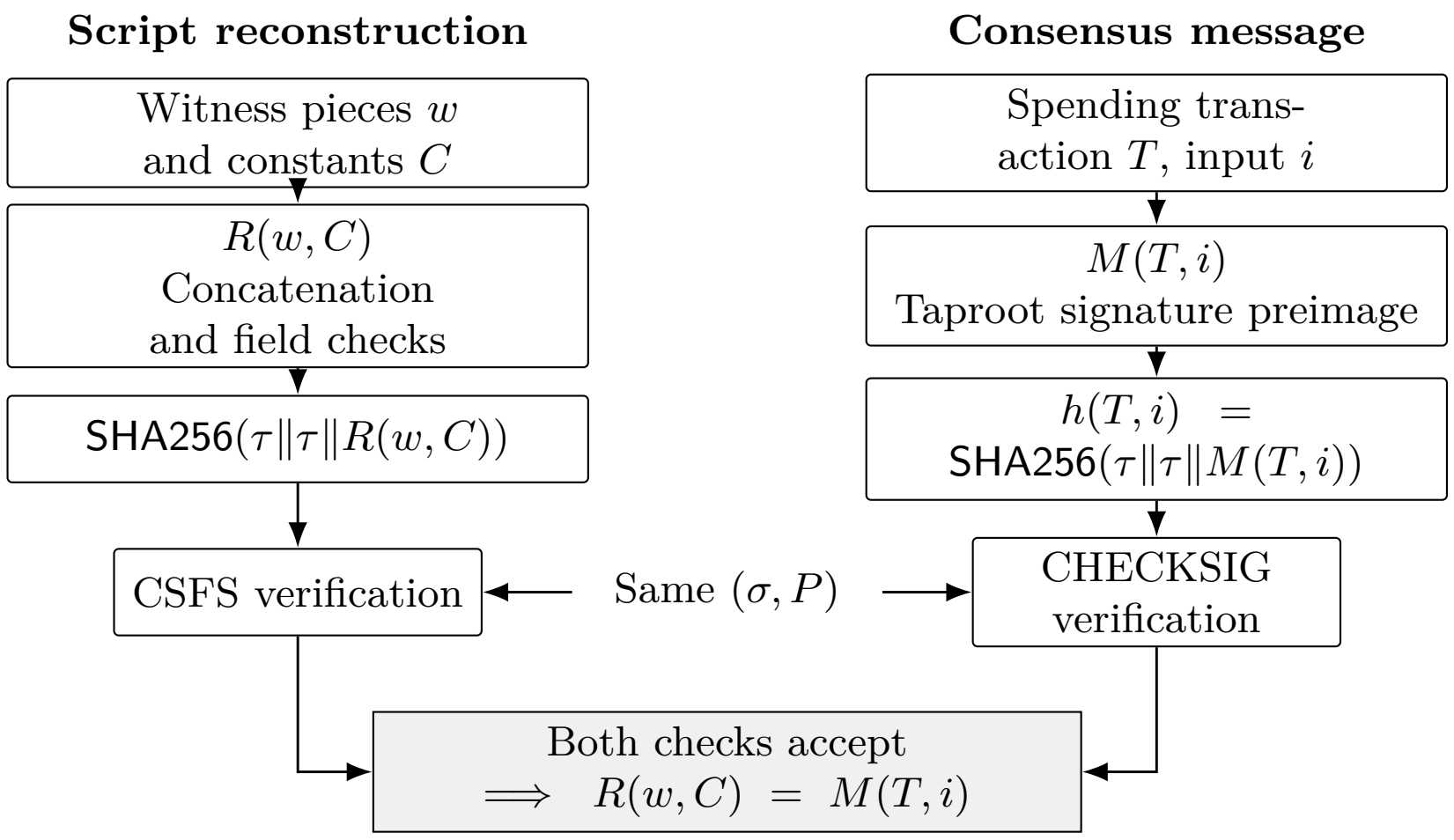

Figure~\ref{fig:same-signature} shows the two messages checked by this single signature. The left digest is reconstructed from witness data; CHECKSIG independently checks the digest determined by the spending transaction.

\begin{lemma}[Binding with a known signing secret]
\label{lem:binding}
Under the stated hash assumptions, acceptance of Eq.~\ref{eq:binding} implies $R=M(T,i)$ except with negligible probability, even if the adversary knows $a$ and chooses $\sigma_a$ adaptively.
\end{lemma}
\begin{proof}
Write the signature as $(r,s)$. BIP-340 verification lifts $r$ to its even-$y$ curve point $R_r$ and checks $sG=R_r+eQ$, where $e$ is the challenge scalar derived from $(r,Q,\mathrm{message})$. Two successful verifications with the same $(r,s,Q)$ give $(e_1-e_2)Q=\mathcal O$. Since $Q$ has prime order, $e_1=e_2$ modulo that order. Distinct messages satisfying this equality violate the assumed challenge-collision resistance. Thus the two 32-byte messages agree. Collision resistance of the tagged TapSighash construction then gives $R=M(T,i)$. Knowing $a$ permits signing each message separately but does not make the same signature valid for two unequal messages.
\end{proof}

\begin{lemma}[Public witness completion]
\label{lem:completion}
For any eligible transaction and tapleaf, a spender who knows the transaction and its spent-output metadata can construct a successful auxiliary signature for a truthful reconstruction without interaction with the protocol signers.
\end{lemma}
\begin{proof}
Compute the consensus preimage, set $R=M(T,i)$, and sign $h(T,i)$ using $a=1$. The two verification arms then receive the same message. Taproot signing does not commit to the auxiliary signature itself. The tapleaf hash commits to the script, whose auxiliary public key is fixed, so signing introduces no self-reference.
\end{proof}

These lemmas separate data authenticity from spending authority. The helper signature is not a secret authorization token. Its role is to ensure that the script checks the actual transaction. The subsequent template check and, where required, an independent authorization signature determine which transaction is permitted.

\section{An Executable CTV-Template Check}
\label{sec:emulation}

\subsection{Witness and reconstruction}

Let $\alpha$ denote an arbitrary lower stack prefix. The macro $E_L$ has the interface
\[
 [\alpha,H,w]\ \longrightarrow\ [\alpha,H],
\]
or fails, where $H$ is the expected 32-byte template hash and $w$ is the auxiliary witness. Its fields are listed in Table~\ref{tab:witness}. Values used more than once are copied with stack operations from one checked item, rather than accepted as unrelated duplicate witness values.

\begin{table}[tb]
\caption{Auxiliary witness fields, after $H$, in bottom-to-top order.}
\label{tab:witness}
\centering\small
\begin{tabular*}{0.9\linewidth}{@{\extracolsep{\fill}}lll}
\toprule
Item & Bytes & Meaning \\
\midrule
$v,\ell$ & $4,4$ & Version and locktime \\
$x$ & 36 & Single input outpoint \\
$A,B$ & $32,32$ & Amount and scriptPubKey aggregate hashes \\
$q$ & 4 & Single input sequence \\
$t$ & 32 & Executing tapleaf hash \\
$(u_j,b_j)_{j=1}^{m}$ & $(8,l_j)$ & Output value and script bytes \\
$\sigma_a$ & 64 & Auxiliary signature under $Q=G$ \\
\bottomrule
\end{tabular*}
\end{table}

The macro checks every length, including $|H|=32$, and computes
\[
\begin{gathered}
 X=\sha(x),\qquad S=\sha(q),\\
 O=\sha\left(\mathop{\|}_{j=1}^{m}
        (u_j\|\operatorname{CompactSize}(l_j)\|b_j)\right).
\end{gathered}
\]
It constructs the preimage shown in Table~\ref{tab:layout}. The four-byte input index, epoch, hash type, spend type, key version and code-separator position are inserted by the script. The other fields are obtained from checked witness items or their hashes.

\begin{table}[tb]
\caption{Exact reconstruction. Offsets are zero-based.}
\label{tab:layout}
\centering\small
\begin{tabular*}{0.8\linewidth}{@{\extracolsep{\fill}}rll}
\toprule
Offset & Bytes & Field \\
\midrule
0 & 2 & Epoch $00$, hash type $00$ \\
2 & 8 & $v\|\ell$ \\
10 & 32 & $X$ (prevouts) \\
42 & 32 & $A$ (amounts) \\
74 & 32 & $B$ (scriptPubKeys) \\
106 & 32 & $S$ (sequences) \\
138 & 32 & $O$ (outputs) \\
170 & 5 & Spend type $02$, input index $00000000$ \\
175 & 32 & $t$ (tapleaf hash) \\
207 & 5 & Key version $00$, separator position $ffffffff$ \\
\bottomrule
\end{tabular*}
\end{table}

The tagged hash input is 276 bytes, within the CAT limit. After the two auxiliary verifications, the macro computes
\begin{equation}
\begin{split}
 C=\sha(&v\|\ell\|\operatorname{le}_4(1)\|S\\
        &\|\operatorname{le}_4(m)\|O\|\operatorname{le}_4(0))
\end{split}
\label{eq:emulated}
\end{equation}
and requires $C=H$. The CTV preimage is 84 bytes. It finally drops the auxiliary items and derived values, leaving $\alpha$ and the original $H$ intact. It does not use the altstack. The supplied compiler emits literal push, PICK, SIZE, EQUALVERIFY, CAT, SHA256, CSFS, VERIFY, CHECKSIGVERIFY and DROP operations; $E_L$ is notation for their expansion, not a new opcode.

\subsection{Soundness, completeness and domain enforcement}

\begin{theorem}[Bounded CTV-template emulation]
\label{thm:ctv}
Consider an otherwise consensus-valid native Taproot script-path spend through the compiled macro, with no executed CODESEPARATOR. Except with negligible probability, $E_L$ accepts a witness only if $T\in\mathcal D_L$ and $H=D(T,0)$. Conversely, for every $T\in\mathcal D_L$ with $H=D(T,0)$, an accepting auxiliary witness can be constructed publicly. On acceptance the macro preserves the lower stack and the original $H$.
\end{theorem}
\begin{proof}
By Lemma~\ref{lem:binding}, the reconstructed 212-byte preimage equals the actual Taproot preimage. Its script-inserted bytes therefore select DEFAULT, no annex, input index zero and the expected extension. All field boundaries are fixed by individual length checks.

The field $X$ equals the actual hash of the concatenated outpoints. Because $x$ is exactly 36 bytes, a different number of actual inputs would require a hash collision between $x$ and a different-length outpoint list. Therefore there is exactly one input. Similarly $S=\sha(q)$ binds its sequence. The current native Taproot input has an empty scriptSig under the assumed consensus-valid spending context; since it is the only input, the additional BIP-119 scriptSig hash is absent.

The field $O$ equals the actual output hash. Hash binding equates the reconstructed output serialization with the actual serialization. Canonical CTxOut serialization is self-delimiting: parsing an eight-byte value, a CompactSize length and the indicated script gives a unique next boundary. Induction over the reconstructed serialization therefore gives exactly $m$ outputs, with script lengths $L$, and the same values and script bytes. Consequently all terms of Eq.~\ref{eq:emulated} are precisely the terms of Eq.~\ref{eq:ctv}; the final equality enforces $H=D(T,0)$.

For completeness, supply the truthful fields and outputs of $T$, and use Lemma~\ref{lem:completion} for $\sigma_a$. Every length and hash check succeeds. The output serialization has length $B_L\leq520$; the other constructed hash inputs have length at most 276. The straight-line stack program copies only its own named fields and removes all temporary items, preserving $\alpha,H$. Resource headroom for an enclosing script remains an explicit composition requirement.
\end{proof}

Thus, on $\mathcal D_L$, existential acceptance of the auxiliary witness is exactly the 32-byte CTV template predicate. Outside this family the emulator can reject transactions that native CTV would accept. It is also not an emulation of non-32-byte NOP behavior. These distinctions prevent a claim of unrestricted opcode equivalence.

\subsection{Fixed templates and signed runtime templates}

A fixed covenant may obtain $H$ from a script constant; it needs no secret authorization signature. A protocol can instead use a runtime $H$ signed by participants. With protocol key $P$ and signature $\sigma_P$, the expanded script has the conceptual form
\[
 E_L\quad \langle P\rangle\quad\csfs,
\]
with entry stack $[\sigma_P,H,w]$. The same $H$ survives $E_L$ and is consumed by CSFS. There is no independently supplied second template hash that could escape the participant's authorization.

\begin{corollary}[Authorization reuse across previous outputs]
\label{cor:reuse}
Let $T,T'\in\mathcal D_L$ have $D(T,0)=D(T',0)=H$, and let both spending policies accept the same protocol key and authorization $\sigma_P$ for $H$. A valid $\sigma_P$ for $T$ remains valid for $T'$; only the public auxiliary witness need be recomputed.
\end{corollary}
\begin{proof}
The protocol CSFS check receives $(\sigma_P,H,P)$ in both cases. By completeness of Theorem~\ref{thm:ctv}, each transaction has a publicly constructible auxiliary witness. Changing an outpoint, spent amount, spent scriptPubKey or executing tapleaf can change $h(T,0)$ and hence $\sigma_a$, without changing $H$ or $\sigma_P$.
\end{proof}

\begin{figure}[tb]
\centering
\begin{tikzpicture}[
  box/.style={draw,rounded corners=2pt,align=center,text width=4.3cm,
              minimum height=1.1cm,inner sep=5pt,font=\small},
  edge/.style={-{Stealth[length=2mm]},thick}]
\node[box] (auth) at (0,0)
  {Participants authorize $H$\\with secret-key signature $\sigma_P$};
\node[box] (helper) at (5.1,0)
  {Spender signs actual $h(T,0)$\\with public helper secret $a=1$};
\node[box] (csfs) at (0,-1.8)
  {CSFS checks $\sigma_P$ for $H$\\under protocol key $P$};
\node[box] (emu) at (5.1,-1.8)
  {$E_L$: bind the reconstruction\\and require $D(T,0)=H$};
\draw[edge] (auth) -- (csfs);
\draw[edge] (helper) -- (emu);
\draw[edge] (emu.west) -- node[above,font=\scriptsize]{$H$} (csfs.east);
\node[align=center,font=\small] at (2.55,-3.05)
  {On an eligible rebind: retain $\sigma_P$; refresh $\sigma_a$ and reconstruction.};
\end{tikzpicture}
\caption{Separate authorization and auxiliary signing. A template authorization remains reusable when the helper witness changes.}
\label{fig:separation}
\end{figure}
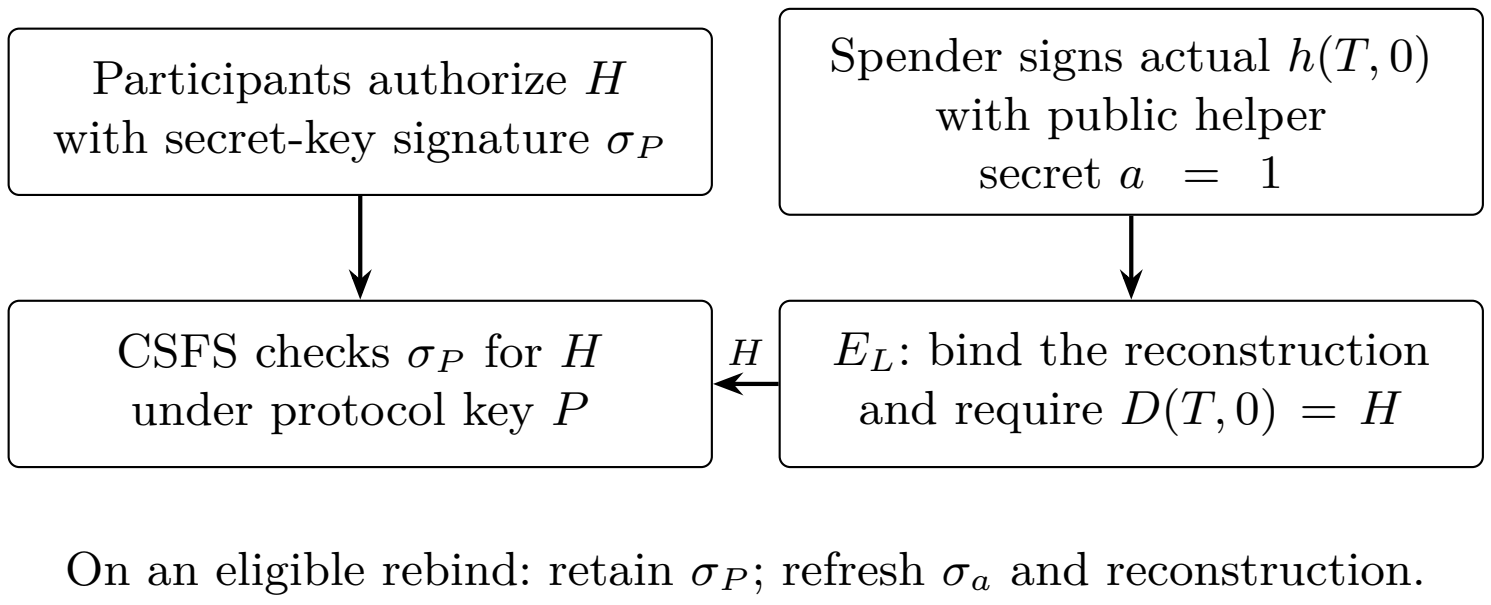

The auxiliary signature itself remains transaction-bound. The corollary concerns a \emph{different} signature, the protocol authorization. Exposing $a$ does not expose the secret for $P$; a valid auxiliary signature is not sufficient to authorize a template. If a deployment assigns both roles to a single secret authorization key, it loses this public-refresh property.

\section{Composition with Lightning Symmetry}
\label{sec:composition}

\subsection{The CTV+CSFS building block}

BIP-348 describes the native block $\mathsf{CTV}\ \langle P\rangle\ \mathsf{CSFS}$ with stack $[\sigma_P,H]$ and relates it to an ALL|ANYPREVOUTANYSCRIPT authorization~\cite{bip348}. CTV authenticates the current transaction against $H$, leaves $H$ on the stack, and CSFS verifies the authorization for that same value. This block supplies reusable template authorization; a complete Lightning Symmetry protocol also supplies state ordering, update and settlement paths, dispute delays and data availability~\cite{decker2018eltoo,rubin2024rekeying}.

Replacing native CTV by $E_L$ preserves that local interface. Importantly, it does not require participants to create a DEFAULT signature whenever an old state is published: the spender creates the DEFAULT auxiliary signature with the public helper secret. This is the precise sense in which retained CHECKSIG commitments do not obstruct protocol-level rebinding.

\subsection{A conditional substitution theorem}

\begin{theorem}[Template-check substitution]
\label{thm:substitution}
Consider a CTV+CSFS protocol whose unilateral transition paths satisfy the following conditions:
\begin{enumerate}
\item each CTV check uses a 32-byte argument and an eligible profile $L$; its allowed transactions lie in $\mathcal D_L$;
\item the remaining script and authorization policy operate on the preserved stack interface, with no executed CODESEPARATOR, annex dependency, or participant signature that newly binds the concrete previous outpoint;
\item the translated transaction templates are built using the translated scripts and then authorized; the UTXO paths enforce the intended policy;
\item each translated spend has sufficient stack, signature-budget, transaction-weight and fee headroom, and required spent-output metadata and protocol data are available.
\end{enumerate}
Replacing each such template check with $E_L$ preserves its transition predicate, up to the addition of a publicly computable auxiliary witness and negligible hash-failure probability.
\end{theorem}
\begin{proof}
For a translated accepting spend, Theorem~\ref{thm:ctv} gives the same template predicate as native CTV and the same stack at the next instruction. All following checks therefore receive their original inputs, including $H$ and any participant authorization. Conversely, for an eligible spend satisfying those checks, completeness supplies the extra witness. Its construction does not require a fresh participant authorization. The compiler uses no altstack space and preserves the lower stack, so the argument applies to each eligible occurrence. Resource and data assumptions ensure the translated execution can be supplied and executed.
\end{proof}

Condition 3 is essential: replacing an output script changes its Taproot address, downstream templates and transaction identifiers. One must construct the translated protocol's templates \emph{before} obtaining state signatures. The local equivalence compares native CTV and $E_L$ on the same translated transaction and template hash; it does not assert byte equality between original and translated transaction graphs. The theorem does not let an existing native-CTV channel keep every old signature while its scripts are rewritten. Within the already translated protocol, Corollary~\ref{cor:reuse} then permits the required authorization reuse across compatible earlier state outputs.

\begin{corollary}[Conditional state-replacement preservation]
If an otherwise sound Lightning Symmetry protocol satisfies Theorem~\ref{thm:substitution}'s conditions, including an enforceable ordering of states and a response delay, then translating its eligible CTV checks preserves the ability to apply an already authorized newer state to a compatible older state output. No APO or native CTV operation is needed in those translated checks.
\end{corollary}
\begin{proof}
Apply the substitution theorem to the newer-state update spending the older output. Its participant signature is unchanged, its helper witness is publicly refreshed, and the preserved surrounding checks still enforce ordering and delay. Apply the same reasoning to subsequent eligible transitions.
\end{proof}

This is a reduction to a specified source protocol, not a proof that every CTV+CSFS channel meets the premises. It does not by itself supply a fee-bumping strategy, watchtower, data-recovery procedure, signing protocol or liveness proof under congestion. In particular, single-input eligibility constrains fee sponsorship; the extra witness changes fees. These issues cannot be inferred away from equality of transaction predicates. Section~\ref{sec:prototype} supplies a concrete bounded prototype with public data recovery and finite fee reserves, and tests its update and settlement paths under full-node rules.

\subsection{Binding the authorized value}
\label{sec:value-binding}

The stack interface matters beyond the CTV comparison itself. A flat sequence of CSFS checks can accept a signed terminal hash $H_1$ while a separate witness item supplies $H_2$ to CTV. If no instruction equates the two items, the signatures do not authorize the checked template. Our local suite executes this counterexample with actual Schnorr signatures. In the present wrapper, $E_L$ preserves the single checked $H$, and the subsequent CSFS consumes that same item. This linkage is necessary for template authorization; state replacement additionally requires the ordering, delay and recovery mechanisms specified above.

\section{Experimental Evaluation}
\label{sec:evaluation}

\newcommand{\NodeCheckRows}{
Template & regtest & 39 & 4 & 35 & 5 \\
Template & Signet & 39 & 4 & 35 & 5 \\
State & regtest & 73 & 14 & 59 & 7 \\
State & Signet & 61 & 12 & 49 & 7 \\
}
\newcommand{\StatePathRows}{
$S_0\to S_1$ & 105/105 & 322226/322226 & 241/496 & 600 \\
$S_1\to S_3$ & 108/108 & 322227/322227 & 241/496 & 1200 \\
$S_3\to$ payouts & 112/112 & 322231/322232 & 185/434 & 600 \\
}

We evaluate whether the compiled check executes under full-node rules, whether a newer state can spend a confirmed stale-state output with preissued participant signatures, and what serialization cost the substitution adds. The artifact combines deterministic opcode checks with Bitcoin Inquisition regtest and public Signet experiments. Native CTV+CSFS is the control. Table~\ref{tab:node-checks} counts named checks, including deliberately invalid candidates, rather than independent statistical trials.

\subsection{Template enforcement}

The experiments used the pinned Bitcoin Inquisition \texttt{v29.4-inq} release with CAT, CSFS and CTV active~\cite{bitcoininquisition}. Covenant outputs use a NUMS internal point and participant keys distinct from the public auxiliary key $Q=G$. Templates for native and emulated channels were constructed before funding. The implementation and saved evidence are available at \url{https://anonymous.4open.science/r/CSFSCAT-8504/}.

\begin{table}[H]
\caption{Recorded full-node checks. Accept denotes mempool acceptance; rejection includes script, policy and missing-input failures. Confirmed counts include funding.}
\label{tab:node-checks}
\centering\small
\begin{tabular*}{\linewidth}{@{\extracolsep{\fill}}llrrrr}
\toprule
Stage & Network & Checks & Accept & Reject & Confirmed \\
\midrule
\NodeCheckRows
\bottomrule
\end{tabular*}
\end{table}

The template stage uses the authorized two-output profile $L=(34,34)$. Each network accepted four valid spends and rejected 34 invalid spends by script verification. The valid cases comprise a baseline emulation, rebinding to a different outpoint and input amount, rebinding to a different tapleaf, and a native control. All four retain the same participant signature on $H$; the three emulated spends have distinct auxiliary signatures. Invalid cases change reconstructed fields, transaction commitments, the authorized template or either signature. The auxiliary signature is refreshed after committed-field mutations, so those rejections cannot be attributed solely to a stale helper signature. One additional annex candidate was rejected by relay policy, not by the theorem's script-enforced annex exclusion.

The four valid Signet spends confirmed after funding. Emulated spends measured 343~vB versus 180~vB for the native control. Selected fees differ and do not provide an equal-feerate comparison.

\subsection{Confirmed state replacement}
\label{sec:prototype}

We exercised two bounded two-party channels, one native and one emulated. Each preauthorizes $S_1,S_2,S_3$ from $S_0$. The on-chain path publishes $S_1$, confirms it, spends its output with the preauthorized $S_3$ update while skipping $S_2$, then settles after a four-block relative delay. Each state leaf has an update branch checking target-state CLTV ordering~\cite{bip65}, the update template and both participants' CSFS signatures on the same $H$; its settlement branch checks CSV with block-based relative-locktime semantics~\cite{bip68,bip112} and a fixed settlement template. Substitution replaces the template checks while retaining these conditions. Updates use output profile $(34,58)$, including an OP\_RETURN metadata output; settlements use $(34,34)$. Public channel parameters, that metadata and signed $S_3$ suffice to recover and verify the spent $S_1$ leaf and value. Fee reserves cover only the specified finite path.

\begin{table}[H]
\caption{Confirmed state-replacement paths. Heights and sizes are native/emulated pairs; fees are per channel in test satoshis.}
\label{tab:state-path}
\centering\small
\begin{tabular*}{\linewidth}{@{\extracolsep{\fill}}lrrrr}
\toprule
Transition & Regtest height & Signet height & vB & Fee \\
\midrule
\StatePathRows
\bottomrule
\end{tabular*}
\end{table}

Both implementations completed the path on both networks (Table~\ref{tab:state-path}). The $S_1\to S_3$ response reused the exact two preissued $S_3$ participant signatures in each channel. The emulated response refreshed the auxiliary signature; replaying one prepared for $S_0\to S_3$ failed. Each settlement paid Alice 3,000 and Bob 2,000 test satoshis. Signet confirmed seven transactions across the channels. Its latest states confirmed at height 322227 and settled at 322231 and 322232, satisfying the four-block delay; the extra block in one path does not establish a performance difference.

The state-protocol run passed 73 named checks on regtest and 61 on Signet. Rejections cover altered authorizations, transaction fields, outputs, metadata and premature settlement. Overfunded regtest fixtures expose backward and same-state CLTV failures without an insufficient-value confound. Missing-input failures after an output was spent are distinguished from script-level ordering failures. These cases test the bounded path, not arbitrary state graphs.

\subsection{Cost and evidence limits}

\newcommand{\ProofTestCount}{170}
\newcommand{\ProofCostRows}{
$(34)$ & 355 & 751 & 283 & 137 \\
$(34,34)$ & 380 & 820 & 343 & 180 \\
$(34,34,34)$ & 407 & 891 & 404 & 223 \\
$(34,83)$ & 380 & 869 & 404 & 229 \\
$34\times12$ & 656 & 1536 & 952 & 610 \\
}

\begin{table}[H]
\caption{Deterministic authorized template-check costs, with one participant signature and a 33-byte control block. Script and witness columns are for the emulator.}
\label{tab:cost}
\centering\small
\begin{tabular*}{\linewidth}{@{\extracolsep{\fill}}lrrrr}
\toprule
Output lengths & Script B & Witness B & vB & Native vB \\
\midrule
\ProofCostRows
\bottomrule
\end{tabular*}
\end{table}

The interpreter passes \ProofTestCount{} named checks across eight output-length profiles, including the 520-byte boundary and rejection of an over-limit profile. Cases cover BIP-340 vector 0~\cite{bip340}, witness and transaction mutations, authorization reuse, counts, annex handling and signature lengths. Five protocol unit tests cover recovery and metadata binding. Local checks complement full-node execution; they do not simulate chain validation or relay policy.

Virtual bytes follow BIP-141's ceiling of transaction weight divided by four~\cite{bip141}. Native and emulated state leaves are 117 and 807 bytes. Updates measure 241 versus 496~vB, an increase of 255~vB (105.8\%); settlements measure 185 versus 434~vB, an increase of 249~vB (134.6\%). The full leaf is supplied on every script-path spend even when only one branch executes. These are deterministic sizes, not CPU or throughput benchmarks. At the selected equal fees, emulated transactions pay a lower feerate, so confirmations do not establish congestion-resistant liveness.

The artifact retains transactions, node responses, block headers and txid inclusion proofs. Offline verifiers recompute template hashes, Taproot commitments, signatures, fees and recorded inclusion relationships. They share code with the producer, do not authenticate witness bytes through the block witness commitment, and do not independently validate the complete Signet chain. Invalid candidates were tested with \texttt{testmempoolaccept}, not mined. The prototype omits HTLCs, routing, automated reorganization recovery and adversarial fee pressure. The evidence supports the stated transaction check and bounded state-replacement execution, not production channel security.

\section{Discussion and Conclusion}
\label{sec:conclusion}

Public auxiliary signing separates transaction binding from protocol authorization. Security depends on keeping the protocol secret separate and preventing bypass paths, including an unrestricted key path controlled by the helper secret.

The substitution result proves the complete BIP-119 template predicate for the stated family, not merely output equality. Its domain describes the construction evaluated here, not a maximal expressiveness result for CSFS+CAT. Supporting larger serialized output vectors, multiple inputs, or variable output-script lengths requires extending the reconstruction and domain checks and accounting for the resulting script, witness and execution costs. We leave the construction, proof and measurement of those extensions outside this paper. For the present implementation, domain restrictions, costs and data availability remain relevant to channel composition: the protocol reduction preserves eligible transition predicates, without asserting equal fees, equal relay behavior or unconditional liveness.

The full-node experiments demonstrate confirmed stale-state override and delayed settlement in a two-party, three-update prototype. They also quantify the cost of the substitution: 255 additional vB per update and 249 per settlement for the measured state scripts. Extending this result requires a practical fee-bumping mechanism, adversarial fee-pressure evaluation, reorganization recovery, and a broader protocol security analysis.

In conclusion, CSFS+CAT can replace native CTV's 32-byte template check on an explicit single-input, bounded-output family. The replacement has soundness and public completeness, and preserves reusable template authorization when the helper and authorization roles are separated. It therefore supports a conditional route to Lightning Symmetry without native CTV or APO. The supplied evidence combines opcode tests, a linkage counterexample, and full-node regtest and Signet executions of the new template and state protocols. It establishes feasibility for the measured bounded paths, while production channel security, unbounded updates and unrestricted equivalence among the proposals remain outside the result.

\bibliographystyle{splncs04}
\ifanonymous
\bibliography{references,references-anonymous}
\else
\bibliography{references,references-author}
\fi

\clearpage
\appendix
\renewcommand{\theHsection}{appendix.\Alph{section}}
\section{Compiler Interface}
\label{app:compiler}

The following pseudocode summarizes the macro, with all named items checked to have the lengths in Table~\ref{tab:witness}. Each use reads a stack copy of that same checked item. Concatenations are in the order written.
\begin{lstlisting}[basicstyle=\ttfamily\small,frame=single]
E_L(alpha, H, v, lock, x, A, B, q, t,
    (value_j, script_j) for j=1..m, aux_sig):
    check all field lengths
    X := SHA256(x)
    S := SHA256(q)
    O := SHA256(concat(value_j, CompactSize(l_j),
                       script_j) for j=1..m)
    M := 0000 || v || lock || X || A || B || S || O
         || 02 || LE32(0) || t || 00 || ffffffff
    h := SHA256(tag || tag || M)
    require CSFS(aux_sig, h, G)
    require CHECKSIG(aux_sig, G)
    C := SHA256(v || lock || LE32(1) || S || LE32(m)
                || O || LE32(0))
    require C == H
    drop all helper fields and derived items
    return alpha, H
\end{lstlisting}

The executable implementation emits the stack-preserving macro used in the protocol experiment. Its local interpreter enforces the emitted subset's stack, CAT-length, cleanstack and signature-budget checks. Its CHECKSIG path obtains a freshly calculated consensus-format digest rather than the script-provided digest; the native CTV control obtains its template from the transaction object rather than witness reconstruction. This separation exposes mismatches between the two calculations, while shared code remains a limitation of the validation. The companion repository contains the implementation, saved transactions and node responses, and detailed reproduction instructions.

\end{document}